\documentstyle[prd,preprint,tighten,aps,eqsecnum,amssymb,amsbsy]{revtex}
\begin{document}
\draft
\preprint{
\begin{tabular}{r}
DFTT 12/95
\\
hep-ph/9502263
\end{tabular}
}
\title{ARE THERE STERILE NEUTRINOS
\\
IN THE FLUX OF SOLAR NEUTRINOS ON THE EARTH?}
\author{S.M. Bilenky\thanks{E-mail address: BILENKY@TO.INFN.IT.}}
\address{
INFN Torino, Via P. Giuria 1, 10125 Torino, Italy
\\
Dipartimento di Fisica Teorica, Universit\`a di Torino
\\
and
\\
Joint Institute for Nuclear Research, Dubna, Russia
}
\author{C. Giunti\thanks{E-mail address: GIUNTI@TO.INFN.IT.}}
\address{
INFN Torino, Via P. Giuria 1, 10125 Torino, Italy
\\
Dipartimento di Fisica Teorica, Universit\`a di Torino
}
\date{February 1994}
\maketitle
\begin{abstract}
It is shown that
the future SNO and Super-Kamiokande experiments,
in which high energy $^8\mathrm{B}$ neutrinos
will be detected through the observation of
CC, NC and $\nu$--$e$ elastic scattering processes,
could allow to reveal in a model independent way
the presence of sterile neutrinos
in the flux of solar neutrinos on the earth.
Lower bounds
for different averaged values
of the probability of transition
of solar $\nu_e$'s into sterile states
and
for the total flux of $^8\mathrm{B}$ neutrinos
are derived in terms of
measurable quantities.
The possibilities to reveal the presence
of $\nu_\mu$ and/or $\nu_\tau$
in the solar neutrino flux on the earth
are also considered
and
the case of transitions
of solar $\nu_e$'s only into sterile states
is discussed.
Some numerical results
for a simple model with
$\nu_e$--$\nu_s$ mixing
are given.
\end{abstract}


\narrowtext

\section{Introduction}

The problem of neutrino masses and mixing
is the most actual one in present day neutrino physics.
As it was first pointed out
by B. Pontecorvo
\cite{B:PONTECORVO},
solar neutrino experiments
are very important tools
for the investigation of this problem.
At present four solar neutrino experiments
(Homestake
\cite{B:HOMESTAKE},
Kamiokande
\cite{B:KAMIOKANDE},
GALLEX
\cite{B:GALLEX}
and
SAGE
\cite{B:SAGE})
investigate different ranges of the solar neutrino spectrum.
The observed event rates in all four experiments
are significantly smaller than
the event rates predicted by the
Standard Solar Model (SSM)
\cite{B:BAHCALL,B:SACLAY,B:CDF}.
Phenomenological analyses
of the existing data
\cite{B:PHENOMEN}
give some indications in favor of
a non-astrophysical explanation
of the observed "deficit"
of solar neutrinos
and it was shown
\cite{B:SOLMSW}
that all data are well described
in the simplest case
of mixing between
$\nu_e$ and $\nu_{\mu}$ (or $\nu_{\tau}$)
if MSW resonant transitions in matter
\cite{B:MSW}
take place.
For the two parameters
$ \Delta m^2 $ and $ \sin^2 2 \theta $
($ \Delta m^2 \equiv m_2^2 - m_1^2 $,
$m_1$ and $m_2$ are the neutrino masses,
and $\theta$ is the mixing angle)
the following values were obtained:
\begin{eqnarray}
\Delta m^2 \simeq 5 \times 10^{-6} \, \mathrm{eV}^2
\quad
\mbox{and}
&
\quad
\sin^2 2\theta \simeq 8 \times 10^{-3}
\quad
&
\mbox{(small mixing angle solution)}
\;,
\label{E:SMAMSW}
\\
\Delta m^2 \simeq 10^{-5} \, \mathrm{eV}^2
\quad
\mbox{and}
&
\quad
\sin^2 2\theta \simeq 0.8
\quad
&
\mbox{(large mixing angle solution)}
\;.
\label{E:LMAMSW}
\end{eqnarray}

The data can also be described by
$\nu_e$ and $\nu_{\mu}$ (or $\nu_{\tau}$)
vacuum oscillations
\cite{B:SOLVAC}
if the parameter
$ \Delta m^2 $
has such a small value that the term
$ \cos \left( \Delta m^2 R \over 2 E \right) $
($R$ is the sun-earth distance
and $E$ is the neutrino energy)
in the expression
for the probability of vacuum transitions
does not disappear due to averaging
(i.e. the oscillation length
is of the same order of magnitude
as the distance between the sun and the earth).
In this case,
for the two parameters
$ \Delta m^2 $ and $ \sin^2 2 \theta $
the following values were found:
\begin{equation}
\Delta m^2 \simeq 8 \times 10^{-11} \, \mathrm{eV}^2
\quad \mbox{and} \quad
\sin^2 2\theta \simeq 0.8
\;.
\label{E:VAC}
\end{equation}

Finally,
all the existing data
can also be described
\cite{B:SOLSTE}
by MSW transitions of solar $\nu_e$'s
into sterile states
(see, for example Ref.\cite{B:BILENKYPETCOV87}).
In this case only the
small mixing angle solution
is allowed\footnote{
Let us stress that
the solutions
(\ref{E:SMAMSW})--(\ref{E:SMASTE})
were obtained under the assumption that
the values of the neutrino fluxes
are given by the SSM.
}:
\begin{equation}
\Delta m^2 \simeq 5 \times 10^{-6} \, \mathrm{eV}^2
\quad \mbox{and} \quad
\sin^2 2\theta \simeq 7 \times 10^{-3}
\quad
\mbox{(small mixing angle solution)}
\;.
\label{E:SMASTE}
\end{equation}
This solution is compatible
with the constraints on the
$\nu_e$--$\nu_s$
mixing parameters obtained
from big bang nucleosynthesis
\cite{B:BBN}.

Thus,
even if we assume that the SSM correctly predicts
the values of the neutrino fluxes from all the main reactions,
the existing solar neutrino data
do not allow to determine in which neutrino states
(active or sterile)
solar $\nu_e$'s are transformed.

The knowledge of the neutrino states
in which solar neutrinos are transferred
is of fundamental importance
for the theory.
Transitions of solar $\nu_e$'s
into active $\nu_\mu$ and/or $\nu_\tau$
are possible in many models beyond the Standard Model
(see, for example, Ref.\cite{B:LANGACKER}).
Such transitions are also possible in the
Standard Model if the neutrino masses and mixing
are generated by the same standard Higgs
mechanism which generates the masses
and mixing of quarks
and the masses of the charged leptons.
On the other hand,
transitions of solar $\nu_e$'s
into sterile states
are possible
only in the models beyond the Standard Model.
Thus,
a discovery of
$ \nu_e \to \nu_{s} $
transitions would be a clear signal
of new physics beyond the Standard Model.

In Ref.\cite{B:BG94A}
we discussed the possibilities of
the solar neutrino experiments
of the next generation
(SNO~\cite{B:SNO},
Super-Kamiokande~\cite{B:SK}),
in which high energy
$^8\mathrm{B}$ neutrinos
will be detected,
to obtain informations
on the transitions
of solar $\nu_e$'s
into sterile states.
Here we will
derive several relations
which could allow
to reveal the presence of sterile neutrinos
in the solar neutrino flux on the earth
independently on any assumption
about the value of the initial
$^8\mathrm{B}$ neutrino flux.
We will also show that from the data of the
SNO and S-K experiments
it will be possible to obtain
a model independent lower bound
for the initial $^8\mathrm{B}$ $\nu_e$ flux
in the general case of transitions of $\nu_e$'s
into active and sterile neutrinos.
A comparison of this lower bound
with the value of the $^8\mathrm{B}$ neutrino flux
predicted by the SSM
will be a direct experimental test of the model
which does not depend on possible
transitions of solar $\nu_e$'s into other states.
Furthermore,
we will discuss the possibilities of SNO and S-K
to reveal the presence of $\nu_\mu$ and/or $\nu_\tau$
in the solar neutrino flux on the earth
and we will consider
the special case of transitions of solar
$\nu_e$'s only into sterile states.
Finally,
we will present some numerical results
obtained in a simple model
with $\nu_e$--$\nu_s$ mixing.

In the SNO experiment
(scheduled to start in the end of 1995)
solar neutrinos will be detected
through the observation of
{\em three} different processes:

\begin{enumerate}

\item
The CC process
\begin{equation}
\nu_{e} + d \to e^{-} + p + p
\;;
\label{E:CC}
\end{equation}

\item
The NC process
\begin{equation}
\nu + d \to \nu + p + n
\;;
\label{E:NC}
\end{equation}

\item
The CC+NC elastic scattering (ES) process
\begin{equation}
\nu + e^{-} \to \nu + e^{-}
\;.
\label{E:ES}
\end{equation}

\end{enumerate}

In the CC process (\ref{E:CC})
the electron spectrum will be measured
and the flux
of solar $\nu_e$'s on the earth
as a function of neutrino energy $E$,
$\phi_{\nu_e}(E)$,
will be determined
\cite{B:SNO}.
The NC process (\ref{E:NC})
will be detected through the observation of neutrons.
Only the total number of NC events will be measured.

In the S-K experiment
(scheduled to start in 1996)
solar neutrinos will be detected
through the observation of the ES process (\ref{E:ES}).
The event rate is expected to be about 50 times
larger than in the current Kamiokande III experiment
and the spectrum of the recoil electrons
will be measured with high accuracy
\cite{B:SK}.

In both the SNO and S-K experiments,
due to the high energy thresholds
($\simeq 6 \, \mathrm{MeV}$
for the CC process,
$2.2 \, \mathrm{MeV}$
for the NC process
and
$\simeq 5 \, \mathrm{MeV}$
for the ES process),
only neutrinos coming from
$^8\mathrm{B}$ decay
will be detected.
The energy spectrum of the initial $^8\mathrm{B}$ $\nu_e$'s
can be written as
\begin{equation}
\phi_{\mathrm{B}}(E)
=
\Phi_{\mathrm{B}}
\,
X(E)
\;.
\label{E500}
\end{equation}
Here $X(E)$ is a known normalized function
determined by the phase space factor\footnote{
It was shown in Ref.\cite{B:BAHCALL91}
that the distortions of the neutrino spectra are
negligibly small under solar conditions.
}
of the decay
$ ^8\mathrm{B} \to \mbox{} ^8\mathrm{Be} + e^{+} + \nu_{e} $,
(the small corrections due to forbidden transitions
where calculated in Ref.\cite{B:BH86})
and
$\Phi_{\mathrm{B}}$
is the total flux of initial $^8\mathrm{B}$ solar $\nu_{e}$'s.

\section{Lower bound for the
$\boldsymbol{^8\mathrm{B}}$ neutrino flux}

In this section we will show that
from the SNO and S-K data
it will be possible to obtain
a model independent {\em lower bound} for
the $^8\mathrm{B}$ neutrino flux $\Phi_{\mathrm{B}}$
in the general case of transitions
of solar $\nu_e$'s into active and sterile neutrinos.
As we will see in the next section,
the knowledge of a lower bound
for the $^8\mathrm{B}$ neutrino flux
will allow to obtain upper bounds for
different averaged values of the probability
of transition of solar $\nu_e$'s
into
$\nu_e$, $\nu_\mu$ and $\nu_\tau$.
If it will occur that any of these upper bounds
is less than one,
it will mean that there are sterile neutrinos
in the flux of solar neutrinos on the earth.

Let us consider first
the NC process (\ref{E:NC}).
Using the
$\nu_{e}$--$\nu_{\mu}$--$\nu_{\tau}$
universality of NC,
we have
\begin{equation}
\left\langle
\sum_{\ell=e,\mu,\tau} \mathrm{P}_{\nu_{e}\to\nu_{\ell}}
\right\rangle_{\mathrm{NC}}
=
{\displaystyle
N^{\mathrm{NC}}
\over\displaystyle
X_{{\nu}d}^{\mathrm{NC}}
\,
\Phi_{\mathrm{B}}
}
\;.
\label{E503}
\end{equation}
Here
$ N^{\mathrm{NC}} $
is the total NC event rate,
$ \mathrm{P}_{\nu_{e}\to\nu_{\ell}} $
is the probability of transition
of solar $\nu_e$'s
into $\nu_{\ell}$
(with $\ell=e,\mu,\tau$)
and
\begin{equation}
X_{{\nu}d}^{\mathrm{NC}}
\equiv
\int_{E_{\mathrm{th}}^{\mathrm{NC}}}
\sigma_{{\nu}d}^{\mathrm{NC}}(E)
\,
X(E)
\,
{\mathrm{d}} E
\;,
\label{E599}
\end{equation}
where
$ \sigma_{{\nu}d}^{\mathrm{NC}}(E) $
is the cross section for the process
$ \nu \, d \to \nu \, n \, p $
and
$ E_{\mathrm{th}}^{\mathrm{NC}} $
is the threshold neutrino energy.
Using the results of a recent calculation of the cross-section
$ \sigma_{{\nu}d}^{\mathrm{NC}}(E) $
\cite{B:KN93}
we obtained
$ X_{{\nu}d}^{\mathrm{NC}} = 4.72 \times 10^{-43} \,\mathrm{cm}^2 $.
The average probability
$ \displaystyle
\left\langle
\sum_{\ell=e,\mu,\tau} \mathrm{P}_{\nu_{e}\to\nu_{\ell}}
\right\rangle_{\mathrm{NC}}
$
is determined as follows:
\begin{equation}
\left\langle
\sum_{\ell=e,\mu,\tau} \mathrm{P}_{\nu_{e}\to\nu_{\ell}}
\right\rangle_{\mathrm{NC}}
\equiv
{\displaystyle
1
\over\displaystyle
X_{{\nu}d}^{\mathrm{NC}}
}
\int_{E_{\mathrm{th}}^{\mathrm{NC}}}
\sigma_{{\nu}d}^{\mathrm{NC}}(E) \,
X(E)
\sum_{\ell=e,\mu,\tau} \mathrm{P}_{\nu_{e}\to\nu_{\ell}}(E) \,
\mathrm{d} E
\;.
\label{E502}
\end{equation}

Taking into account that
$ \displaystyle
\left\langle
\sum_{\ell=e,\mu,\tau} \mathrm{P}_{\nu_{e}\to\nu_{\ell}}
\right\rangle_{\mathrm{NC}}
\le 1
$,
from Eq.(\ref{E503})
we obtain the following lower bound
for the total flux of $^8\mathrm{B}$ neutrinos:
\begin{equation}
\Phi_{\mathrm{B}}
\ge
{\displaystyle
N^{\mathrm{NC}}
\over\displaystyle
X_{{\nu}d}^{\mathrm{NC}}
}
\;.
\label{E518}
\end{equation}

Another lower bound for the flux $\Phi_{\mathrm{B}}$
can be obtained from the CC data.
The $\nu_e$ survival probability
is connected with the flux of $\nu_e$
on the earth by the relation
\begin{equation}
\mathrm{P}_{\nu_{e}\to\nu_{e}}(E)
=
{\displaystyle
\phi_{\nu_{e}}(E)
\over\displaystyle
X(E)
\,
\Phi_{\mathrm{B}}
}
\;.
\label{E555}
\end{equation}
{}From this relation it follows that
\begin{equation}
\Phi_{\mathrm{B}}
\ge
\left[
{\displaystyle
\phi_{\nu_e}(E)
\over\displaystyle
X(E)
}
\right]_{\mathrm{max}}
\;,
\label{E560}
\end{equation}
where
$ \displaystyle
\left[
\phi_{\nu_e}(E)
/
X(E)
\right]_{\mathrm{max}}
$
is the maximal value
of the function
$ \displaystyle
\phi_{\nu_e}(E)
/
X(E)
$
in the explored energy range.

Analogously,
we have
\begin{equation}
\Phi_{\mathrm{B}}
\ge
{\displaystyle
N^{\mathrm{CC}}
\over\displaystyle
X_{\nu_{e}d}^{\mathrm{CC}}
}
\;.
\label{E809}
\end{equation}
Here
$ N^{\mathrm{CC}} $
is the total CC event rate
and
\begin{equation}
X_{\nu_{e}d}^{\mathrm{CC}}
\equiv
\int_{E_{\mathrm{th}}^{\mathrm{CC}}}
\sigma_{\nu_{e}d}^{\mathrm{CC}}(E)
\,
X(E)
\,
{\mathrm{d}} E
\;,
\label{E810}
\end{equation}
where
$ \sigma_{\nu_{e}d}^{\mathrm{CC}}(E) $
is the cross section for the process
$ \nu_e \, d \to e^{-} \, p \, p $.
Using the results of
Ref.\cite{B:KN93}
we obtained
$ X_{\nu_{e}d}^{\mathrm{CC}} \simeq 1.1 \times 10^{-42} \, \mathrm{cm}^2 $
for
$ E_{\mathrm{th}}^{\mathrm{CC}} \simeq 6 \, \mathrm{MeV} $.

The total number of ES events
$ N^{\mathrm{ES}} $
is given by
\begin{equation}
N^{\mathrm{ES}}
=
\int_{{E_{\mathrm{th}}^{\mathrm{ES}}}}
\sum_{\ell=e,\mu,\tau}
\sigma_{\nu_{\ell}e}(E)
\mathrm{P}_{\nu_{e}\to\nu_{\ell}}(E)
\,
X(E)
\,
{\mathrm{d}} E
\,
\Phi_{\mathrm{B}}
\;,
\label{E506}
\end{equation}
where
$ \sigma_{\nu_\ell e}(E) $
is the total cross section of the process
$ \nu_\ell \, e \to \nu_\ell \, e $
(with $\ell=e,\mu,\tau$).
Taking into account that
$ \sigma_{\nu_\mu e}(E) / \sigma_{\nu_e e}(E)
\simeq 1/6 $,
from Eq.(\ref{E506})
we have
\begin{equation}
N^{\mathrm{ES}}
\le
\int_{{E_{\mathrm{th}}^{\mathrm{ES}}}}
\sigma_{\nu_{e}e}(E)
\sum_{\ell=e,\mu,\tau}
\mathrm{P}_{\nu_{e}\to\nu_{\ell}}(E)
\,
X(E)
\,
{\mathrm{d}} E
\,
\Phi_{\mathrm{B}}
\;,
\label{E801}
\end{equation}
{}From Eq.(\ref{E801})
we obtain the following lower bound for $\Phi_{\mathrm{B}}$
\begin{equation}
\Phi_{\mathrm{B}}
\ge
{\displaystyle
N^{\mathrm{ES}}
\over\displaystyle
X_{\nu_{e}e}
}
\;,
\label{E802}
\end{equation}
where
\begin{equation}
X_{\nu_{e}e}
\equiv
\int_{E_{\mathrm{th}}^{\mathrm{ES}}}
\sigma_{\nu_{e}e}(E)
\,
X(E)
\,
{\mathrm{d}} E
\label{E572}
\end{equation}
For
$ E_{\mathrm{th}}^{\mathrm{ES}} \simeq 5 \, \mathrm{MeV} $
we have
$ \displaystyle
X_{\nu_{e}e}
\simeq
2 \times 10^{-44} \, \mathrm{cm}^2
$.
The values of
$ N^{\mathrm{ES}} $
and
$ X_{\nu_{e}e} $
depend on the threshold energy
$ E_{\mathrm{th}}^{\mathrm{ES}} $.
Let us notice that it is worthwhile
to study the dependence of
$ N^{\mathrm{ES}} / X_{\nu_{e}e} $
on the threshold energy
$ E_{\mathrm{th}}^{\mathrm{ES}} $
in order to choose
the optimal value of this quantity
(providing that the statistical accuracy
is sufficiently high).

\begin{figure}[p]
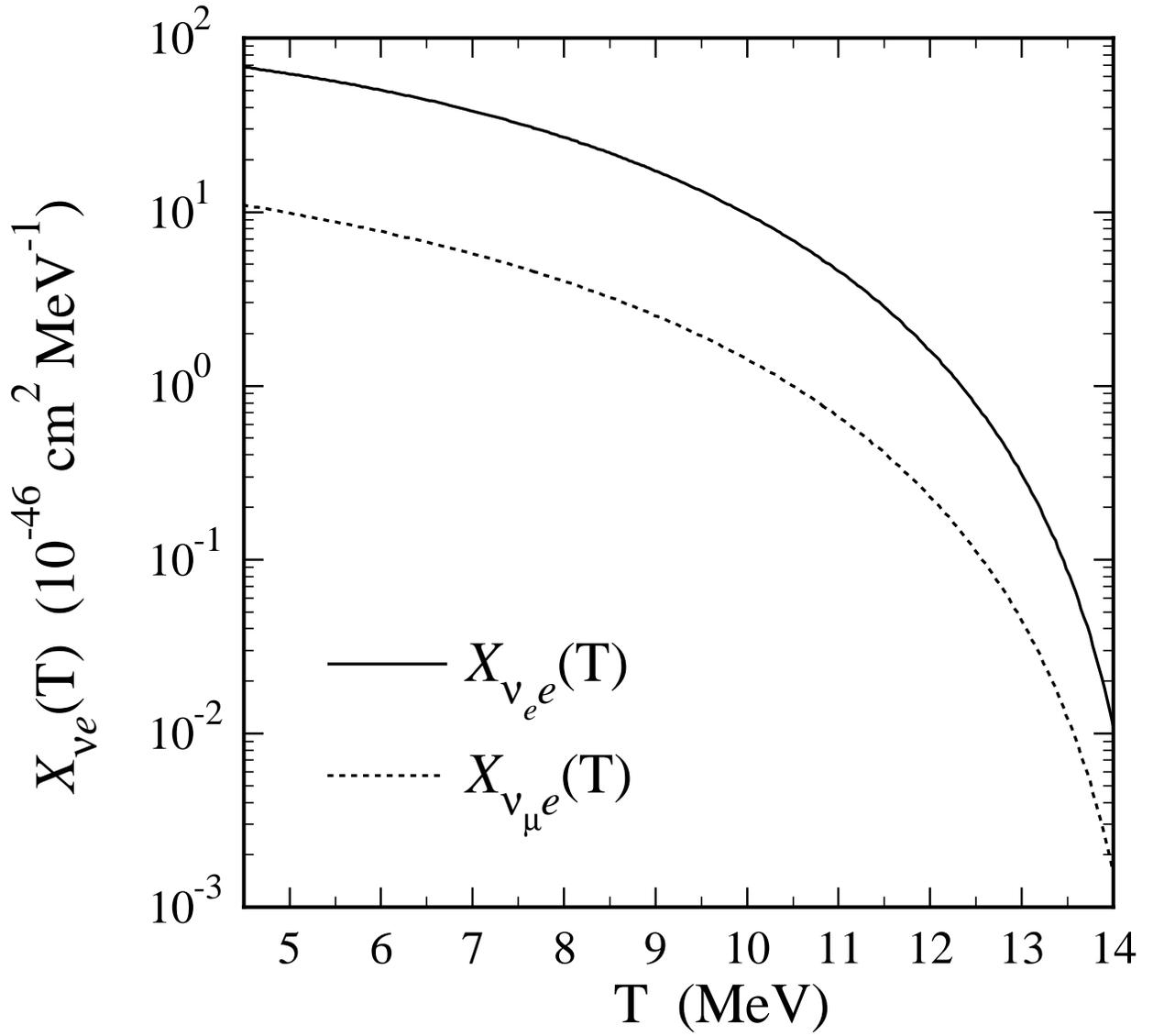

\protect\caption{
Plot of the functions
$ X_{\nu_{\mu}e}(\mathrm{T}) $
and
$ X_{\nu_{e}e}(\mathrm{T}) $
defined in
Eqs.(\protect\ref{E533}) and (\protect\ref{E575}),
respectively.
The depicted range for
the kinetic energy
$\mathrm{T}$
of the recoil electrons in the ES process
will be explored by SNO
with $\mathrm{T}_{\mathrm{th}}=4.5\,\mathrm{MeV}$.}
\label{F:XES}
\end{figure}

In the S-K experiment
the spectrum of ES recoil electrons
$ n^{\mathrm{ES}}(\mathrm{T}) $
($\mathrm{T}$ is the electron kinetic energy)
will be measured with high accuracy.
We have
\begin{equation}
n^{\mathrm{ES}}(\mathrm{T})
=
\int_{E_{\mathrm{m}}(\mathrm{T})}
\sum_{\ell=e,\mu,\tau}
{\displaystyle
\mathrm{d} \sigma_{\nu_{\ell}e}
\over\displaystyle
\mathrm{d} \mathrm{T}
}
(E,\mathrm{T})
\,
\mathrm{P}_{\nu_{e}\to\nu_{\ell}}(E)
\,
X(E)
\,
\mathrm{d} E
\,
\Phi_{\mathrm{B}}
\;.
\label{E531}
\end{equation}
Here
$ \displaystyle
{\displaystyle
\mathrm{d} \sigma_{\nu_{\ell}e}
\over\displaystyle
\mathrm{d} \mathrm{T}
}
(E,\mathrm{T})
$
is the differential cross section
of the process
$ \nu_{\ell} e \to \nu_{\ell} e $
(with $\ell=e,\mu,\tau$)
and
$
E_{\mathrm{m}}(\mathrm{T})
=
{1\over2}
\,
\mathrm{T}
\left(
1
+
\sqrt{ 1 + 2 \, m_{e} / \mathrm{T} }
\right)
$.
{}From Eq.(\ref{E531})
it follows that
\begin{equation}
n^{\mathrm{ES}}(\mathrm{T})
\le
\int_{E_{\mathrm{m}}(\mathrm{T})}
{\displaystyle
\mathrm{d} \sigma_{\nu_{e}e}
\over\displaystyle
\mathrm{d} \mathrm{T}
}
(E,\mathrm{T})
\sum_{\ell=e,\mu,\tau}
\mathrm{P}_{\nu_{e}\to\nu_{\ell}}(E)
\,
X(E)
\,
\mathrm{d} E
\,
\Phi_{\mathrm{B}}
\;.
\label{E803}
\end{equation}
{}From Eq.(\ref{E803})
we obtain the following lower bound
for the total $^8\mathrm{B}$ neutrino flux:
\begin{equation}
\Phi_{\mathrm{B}}
\ge
\left[
{\displaystyle
n^{\mathrm{ES}}(\mathrm{T})
\over\displaystyle
X_{\nu_{e}e}(\mathrm{T})
}
\right]_{\mathrm{max}}
\;,
\label{E804}
\end{equation}
where
\begin{equation}
X_{\nu_{e}e}(\mathrm{T})
\equiv
\int_{E_{\mathrm{m}}(\mathrm{T})}
{\displaystyle
\mathrm{d} \sigma_{\nu_{e}e}
\over\displaystyle
\mathrm{d} \mathrm{T}
}
(E,\mathrm{T})
\,
X(E)
\,
\mathrm{d} E
\;.
\label{E575}
\end{equation}
The function
$ X_{\nu_{e}e}(\mathrm{T}) $
is plotted in Fig.\ref{F:XES}.

It is clear from the derivation of
Eqs.(\ref{E802}) and (\ref{E804})
that,
if an appreciable part of solar $\nu_e$'s
are transformed into $\nu_{\mu}$ and/or $\nu_{\tau}$,
the lower bounds
(\ref{E802}) and (\ref{E804})
could be far away from the true value
of the $^8\mathrm{B}$ neutrino flux.
We will obtain now
additional inequalities which do not have
this drawback.
{}From Eq.(\ref{E506})
we have
\begin{equation}
\left\langle
\sum_{\ell=e,\mu,\tau} \mathrm{P}_{\nu_{e}\to\nu_{\ell}}
\right\rangle_{\mathrm{ES}}
=
{\displaystyle
\Sigma^{\mathrm{ES}}
\over\displaystyle
X_{\nu_{\mu}e}
\,
\Phi_{\mathrm{B}}
}
\;.
\label{E509}
\end{equation}
Here
\begin{equation}
\Sigma^{\mathrm{ES}}
\equiv
N^{\mathrm{ES}}
-
\int_{{E_{\mathrm{th}}^{\mathrm{ES}}}}
\left(
\sigma_{\nu_{e}e}(E)
-
\sigma_{\nu_{\mu}e}(E)
\right)
\phi_{\nu_{e}}(E)
\,
{\mathrm{d}} E
\;,
\label{E508}
\end{equation}
\begin{equation}
\left\langle
\sum_{\ell=e,\mu,\tau} \mathrm{P}_{\nu_{e}\to\nu_{\ell}}
\right\rangle_{\mathrm{ES}}
\equiv
{\displaystyle
1
\over\displaystyle
X_{\nu_{\mu}e}
}
\int_{E_{\mathrm{th}}^{\mathrm{ES}}}
\sigma_{\nu_{\mu}e}(E)
\,
X(E)
\sum_{\ell=e,\mu,\tau} \mathrm{P}_{\nu_{e}\to\nu_{\ell}}(E)
\,
\mathrm{d} E
\;.
\label{E511}
\end{equation}
and
\begin{equation}
X_{\nu_{\mu}e}
\equiv
\int_{E_{\mathrm{th}}^{\mathrm{ES}}}
\sigma_{\nu_{\mu}e}(E)
\,
X(E)
\,
{\mathrm{d}} E
\;.
\label{E510}
\end{equation}
For
$ E_{\mathrm{th}}^{\mathrm{ES}} \simeq 6 \, \mathrm{MeV} $
(which corresponds to a
kinetic energy threshold
$ \mathrm{T}_{\mathrm{th}} = 4.5 \, \mathrm{MeV} $
for the electrons in the CC process)
we have
$ \displaystyle
X_{\nu_{\mu}e}
\simeq
2 \times 10^{-45} \, \mathrm{cm}^2
$.
{}From Eq.(\ref{E509})
we obtain the following lower bound
for the total $^8\mathrm{B}$ neutrino flux:
\begin{equation}
\Phi_{\mathrm{B}}
\ge
{\displaystyle
\Sigma^{\mathrm{ES}}
\over\displaystyle
X_{\nu_{\mu}e}
}
\;.
\label{E524}
\end{equation}
The quantity
$ \Sigma^{\mathrm{ES}} $
can be obtained directly
from the data of the SNO and S-K experiments.
In fact,
$ N^{\mathrm{ES}} $
will be measured in both experiments.
The second term in the right-hand side
of Eq.(\ref{E508})
can be determined from the CC data of the SNO experiment.

Finally,
from Eq.(\ref{E531})
we obtain
\begin{equation}
\left\langle
\sum_{\ell=e,\mu,\tau} \mathrm{P}_{\nu_{e}\to\nu_{\ell}}
\right\rangle_{\mathrm{ES};\mathrm{T}}
=
{\displaystyle
\Sigma^{\mathrm{ES}}(\mathrm{T})
\over\displaystyle
X_{\nu_{\mu}e}(\mathrm{T})
\,
\Phi_{\mathrm{B}}
}
\;.
\label{E532}
\end{equation}
Here
\begin{equation}
\Sigma^{\mathrm{ES}}(\mathrm{T})
\equiv
n^{\mathrm{ES}}(\mathrm{T})
-
\int_{E_{\mathrm{m}}(\mathrm{T})}
\left[
{\displaystyle
\mathrm{d} \sigma_{\nu_{e}e}
\over\displaystyle
\mathrm{d} \mathrm{T}
}
(E,\mathrm{T})
-
{\displaystyle
\mathrm{d} \sigma_{\nu_{\mu}e}
\over\displaystyle
\mathrm{d} \mathrm{T}
}
(E,\mathrm{T})
\right]
\phi_{\nu_{e}}(E)
\,
\mathrm{d} E
\;,
\label{E534}
\end{equation}
\begin{equation}
\left\langle
\sum_{\ell=e,\mu,\tau} \mathrm{P}_{\nu_{e}\to\nu_{\ell}}
\right\rangle_{\mathrm{ES};\mathrm{T}}
\equiv
{\displaystyle
1
\over\displaystyle
X_{\nu_{\mu}e}(\mathrm{T})
}
\int_{E_{\mathrm{m}}(\mathrm{T})}
{\displaystyle
\mathrm{d} \sigma_{\nu_{\mu}e}
\over\displaystyle
\mathrm{d} \mathrm{T}
}
(E,\mathrm{T})
\,
X(E)
\sum_{\ell=e,\mu,\tau} \mathrm{P}_{\nu_{e}\to\nu_{\ell}}(E) \,
\mathrm{d} E
\label{E535}
\end{equation}
and
\begin{equation}
X_{\nu_{\mu}e}(\mathrm{T})
\equiv
\int_{E_{\mathrm{m}}(\mathrm{T})}
{\displaystyle
\mathrm{d} \sigma_{\nu_{\mu}e}
\over\displaystyle
\mathrm{d} \mathrm{T}
}
(E,\mathrm{T})
\,
X(E)
\,
\mathrm{d} E
\;.
\label{E533}
\end{equation}
The function
$ X_{\nu_{\mu}e}(\mathrm{T}) $
is plotted in Fig.\ref{F:XES}.
{}From Eq.(\ref{E532})
we obtain the following lower bound
for the total $^8\mathrm{B}$ neutrino flux:
\begin{equation}
\Phi_{\mathrm{B}}
\ge
\left[
{\displaystyle
\Sigma^{\mathrm{ES}}(\mathrm{T})
\over\displaystyle
X_{\nu_{\mu}e}(\mathrm{T})
}
\right]_{\mathrm{max}}
\;.
\label{E539}
\end{equation}

Thus,
we obtained several expressions
which give a lower bound
for the total flux of $^8\mathrm{B}$ neutrinos
(Eqs.(\ref{E518}), (\ref{E560}), (\ref{E802}),
(\ref{E804}), (\ref{E524}), (\ref{E539})).
All these expressions depend only on
known quantities
and quantities that will be measured in
the SNO and S-K experiments.
When the SNO and S-K data
will be available
it will be possible to choose
the expression which gives the highest lower bound.
This will be the best model independent
lower bound for
the total flux of $^8\mathrm{B}$ neutrinos
in the general case
of transitions of solar $\nu_e$'s
into active and sterile states.
We will denote this best lower bound as
$ \Phi_{\mathrm{B}}^{\circ} $.

A comparison of
$ \Phi_{\mathrm{B}}^{\circ} $
with the value of
the total flux of $^8\mathrm{B}$ neutrinos
predicted by the SSM,
$ \Phi_{\mathrm{B}}^{\mathrm{SSM}} $,
will be a direct experimental test
of the model.
If
$ \Phi_{\mathrm{B}}^{\circ}
\le
\Phi_{\mathrm{B}}^{\mathrm{SSM}} $
no conclusion about the SSM
can be reached.
If
$ \Phi_{\mathrm{B}}^{\circ}
>
\Phi_{\mathrm{B}}^{\mathrm{SSM}} $
it would mean that the
$^8\mathrm{B}$ neutrino flux
is higher than the flux predicted by the SSM.

In the next section
we will use
$ \Phi_{\mathrm{B}}^{\circ} $
in order to obtain several inequalities
whose test will allow
to reveal the presence of sterile neutrinos
in the flux of solar neutrinos on the earth.

\section{Sterile neutrino tests}

In this section
we will obtain several inequalities
which could allow to reveal
the presence of sterile neutrinos in the flux
of solar neutrinos on the earth.
Let us consider Eq.(\ref{E503}).
In the case of transitions of solar $\nu_e$'s
only into
$\nu_e$, $\nu_\mu$ and $\nu_\tau$
we have
$ \displaystyle
\left\langle
\sum_{\ell=e,\mu,\tau} \mathrm{P}_{\nu_{e}\to\nu_{\ell}}
\right\rangle_{\mathrm{NC}}
= 1
$.
In the general case of transitions of $\nu_e$'s
into active and sterile states\footnote{
We assume that neutrinos are stable particles.
For a discussion
of neutrino instability
see Ref.\cite{B:BMA}
and references therein.
}.
\begin{equation}
\left\langle
\sum_{\ell=e,\mu,\tau} \mathrm{P}_{\nu_{e}\to\nu_{\ell}}
\right\rangle_{\mathrm{NC}}
=
1
-
\left\langle
\mathrm{P}_{\nu_{e}\to\nu_{s}}
\right\rangle_{\mathrm{NC}}
< 1
\;.
\label{E825}
\end{equation}
Here
$ \mathrm{P}_{\nu_{e}\to\nu_{s}} $
is the probability of transition of solar $\nu_e$'s
into all possible sterile states.
The right-hand side of Eq.(\ref{E503})
depends on the total flux
of $^8\mathrm{B}$ neutrinos
$ \Phi_{\mathrm{B}} $.
In the case of transitions
of solar $\nu_e$'s
into sterile states the flux
$ \Phi_{\mathrm{B}} $
cannot be determined from the experimental data.
However,
as we have seen in the previous section,
from the SNO and S-K data
it will be possible to determine
a lower bound for the total flux
of $^8\mathrm{B}$ neutrinos,
$ \Phi_{\mathrm{B}}^{\circ} $.
Therefore,
from Eq.(\ref{E503}) we obtain
\begin{equation}
\left\langle
\sum_{\ell=e,\mu,\tau} \mathrm{P}_{\nu_{e}\to\nu_{\ell}}
\right\rangle_{\mathrm{NC}}
\le
{\displaystyle
N^{\mathrm{NC}}
\over\displaystyle
X_{{\nu}d}^{\mathrm{NC}}
\,
\Phi_{\mathrm{B}}^{\circ}
}
\;.
\label{E805}
\end{equation}
The right-hand side of Eq.(\ref{E805})
contains only measurable quantities.
If it will turn out that
$ \displaystyle
N^{\mathrm{NC}}
/
X_{{\nu}d}^{\mathrm{NC}}
\Phi_{\mathrm{B}}^{\circ}
< 1
$,
it will mean that
$ \nu_e \to \nu_{s} $
transitions take place.
{}From Eqs.(\ref{E825}) and (\ref{E805}),
for the averaged value
of the probability
of transitions of $\nu_e$'s
into all possible sterile states
$ \displaystyle
\left\langle
\mathrm{P}_{\nu_{e}\to\nu_{s}}
\right\rangle_{\mathrm{NC}}
$
we find the following lower bound:
\begin{equation}
\left\langle
\mathrm{P}_{\nu_{e}\to\nu_{s}}
\right\rangle_{\mathrm{NC}}
\ge
1
-
{\displaystyle
N^{\mathrm{NC}}
\over\displaystyle
X_{{\nu}d}^{\mathrm{NC}}
\,
\Phi_{\mathrm{B}}^{\circ}
}
\;.
\label{E806}
\end{equation}

Analogously,
from Eqs.(\ref{E509}) and (\ref{E532})
we obtain the following inequalities
for other averages of the probability of transition
of solar $\nu_e$'s
into all possible sterile states\footnote{
It is easy to see that if
$ \mathrm{P}_{\nu_{e}\to\nu_{s}}(E) = \mbox{const} $
the ratios in the right-hand sides
of Eqs.(\ref{E806})--(\ref{E808})
are larger or equal to one.
Therefore,
in this case
one cannot reach any conclusion
about transitions of solar $\nu_e$'s
into sterile states.
}:
\begin{equation}
\left\langle
\mathrm{P}_{\nu_{e}\to\nu_{s}}
\right\rangle_{\mathrm{ES}}
\ge
1
-
{\displaystyle
\Sigma^{\mathrm{ES}}
\over\displaystyle
X_{\nu_{\mu}e}
\,
\Phi_{\mathrm{B}}^{\circ}
}
\;,
\label{E807}
\end{equation}
\begin{equation}
\left\langle
\mathrm{P}_{\nu_{e}\to\nu_{s}}
\right\rangle_{\mathrm{ES};\mathrm{T}}
\ge
1
-
{\displaystyle
\Sigma^{\mathrm{ES}}(\mathrm{T})
\over\displaystyle
X_{\nu_{\mu}e}(\mathrm{T})
\,
\Phi_{\mathrm{B}}^{\circ}
}
\;.
\label{E808}
\end{equation}
If it will turn out that
the right-hand side
of at least one of these inequalities
is different from zero,
it will mean that
active neutrinos transform into sterile states
and that the probability
of $\nu_e\to\nu_s$ transitions
depends on the neutrino energy $E$.
Let us stress again
that a discovery of such transitions
will have a great impact on
the theory.

\section{Transitions of solar
$\boldsymbol{\nu_{\lowercase{e}}}$'s
only into sterile states}

If the tests proposed in the previous section
will demonstrate the presence of
$ \nu_e \to \nu_s $
transitions,
the question will arise
whether there are also
transitions of solar $\nu_e$'s
into active
$\nu_\mu$ and/or $\nu_\tau$.
When solar neutrinos
will be detected through the observation
of CC, NC and ES processes,
it will be possible to reveal
the presence of $\nu_\mu$ and/or $\nu_\tau$
in the flux of solar neutrinos on the earth
independently from
$\nu_e\to\nu_s$ transitions
\cite{B:BG93}.
In fact,
let us consider the quantities
\begin{eqnarray}
&&
R^{\mathrm{ES}}
=
1
-
{\displaystyle
1
\over\displaystyle
N^{\mathrm{ES}}
}
\int_{E_{\mathrm{th}}^{\mathrm{ES}}}
\sigma_{\nu_{e}e}(E)
\,
\phi_{\nu_{e}}(E)
\,
\mathrm{d} E
\;,
\label{E811}
\\
&&
R^{\mathrm{NC}}
=
1
-
{\displaystyle
1
\over\displaystyle
N^{\mathrm{NC}}
}
\int_{E_{\mathrm{th}}^{\mathrm{NC}}}
\sigma_{{\nu}d}^{\mathrm{NC}}(E)
\,
\phi_{\nu_{e}}(E)
\,
\mathrm{d} E
\;,
\label{E812}
\end{eqnarray}
which give the relative contribution of
$\nu_\mu$ and $\nu_\tau$
to the ES and NC event rates,
respectively.
Here
$ \phi_{\nu_{e}}(E) $
is the flux of solar $\nu_e$'s on the earth,
which will be determined for
$ E \ge E_{\mathrm{th}}^{\mathrm{CC}} \simeq 6 \, \mathrm{MeV} $
from the CC data of the SNO experiment.
Since in both the SNO and S-K
experiment it is possible to choose
an energy threshold for the ES process
$ E_{\mathrm{th}}^{\mathrm{ES}} \ge 6 \, \mathrm{MeV} $,
the ratio
$ R^{\mathrm{ES}} $
will be determined directly from the experimental data.
On the other hand,
the NC threshold in the SNO experiment is fixed at
$ E_{\mathrm{th}}^{\mathrm{NC}} = 2.2 \, \mathrm{MeV} $,
but
there will be no direct experimental information
on the value of
$ \phi_{\nu_{e}}(E) $
in the energy interval
$
E_{\mathrm{th}}^{\mathrm{NC}}
\le
E
\le
E_{\mathrm{th}}^{\mathrm{CC}}
$.
It is obvious that
\begin{equation}
R^{\mathrm{NC}}
\le
1
-
{\displaystyle
1
\over\displaystyle
N^{\mathrm{NC}}
}
\int_{E_{\mathrm{th}}^{\mathrm{CC}}}
\sigma_{{\nu}d}^{\mathrm{NC}}(E)
\,
\phi_{\nu_{e}}(E)
\,
\mathrm{d} E
\;.
\label{E820}
\end{equation}
Let us notice that
the contribution of the integral
$ \displaystyle
{\displaystyle
1
\over\displaystyle
N^{\mathrm{NC}}
}
\int_{E_{\mathrm{th}}^{\mathrm{NC}}}^{E_{\mathrm{th}}^{\mathrm{CC}}}
\sigma_{{\nu}d}^{\mathrm{NC}}(E)
\,
\phi_{\nu_{e}}(E)
\,
\mathrm{d} E
$
to the right-hand part of Eq.(\ref{E812})
is expected to be small
due to the smallness of the cross section
in the corresponding energy interval.
Our model calculations show
that this contribution is of the order
of a few percentage.

In the following we assume
that from the ratio
$ R^{\mathrm{ES}} $
and the upper bound (\ref{E820}) for
$ R^{\mathrm{NC}} $
no indication
in favor of the presence of
$\nu_\mu$ and/or $\nu_\tau$
in the solar neutrino flux on the earth
will be found.
If,
in addition,
it will be found that
the spectrum of solar $\nu_e$'s is distorted
(i.e. the quantity
$ \phi_{\nu_e}(E) / X(E) $
is energy dependent),
it will mean that
solar $\nu_e$'s transform
only into sterile states.
These transitions could take place
if $\nu_e$ is mixed only with sterile neutrinos
(as in the models presented in Ref.\cite{B:STEMIX})
or
if neutrinos have large Dirac magnetic moments
($\sim$ $10^{-11}$--$10^{-10}$ $\mu_{B}$)
and left-handed solar $\nu_e$'s
transform partly into right-handed sterile states
due to precessions in the magnetic field of the sun
\cite{B:MAGNETIC}.
As it is well known,
in the last case
the flux of solar $\nu_e$'s
has a time dependence
anti-correlated with the solar activity.

If solar neutrinos are transformed
only into sterile states,
the best lower bound
for the total $^8\mathrm{B}$ neutrino flux
is given by Eq.(\ref{E560}).
In fact,
in this case
$
\left[ \mathrm{P}_{\nu_{e}\to\nu_{e}} \right]_{\mathrm{max}}
\ge
\left\langle
\mathrm{P}_{\nu_{e}\to\nu_{e}}
\right\rangle_{\mathrm{a}}
$
(with $a=\mathrm{NC},\mathrm{ES},\ldots$).
{}From this inequality it follows that
\begin{equation}
\Phi_{\mathrm{B}}^{\circ}
=
\left[
{\displaystyle
\phi_{\nu_e}(E)
\over\displaystyle
X(E)
}
\right]_{\mathrm{max}}
\;.
\label{E830}
\end{equation}

Let us discuss now
what informations
about the probability
of $\nu_e$'s to survive
can be obtained from the experimental data
in the case under consideration.
{}From the measurement of the charged current
spectrum the probability
$ \mathrm{P}_{\nu_{e}\to\nu_{e}}(E) $
can be determined up to a constant
(see Eq.(\ref{E555})).
{}From Eq.(\ref{E830})
we obtain
the following
upper bound for the survival probability:
\begin{equation}
\mathrm{P}_{\nu_{e}\to\nu_{e}}(E)
\le
{\displaystyle
\phi_{\nu_{e}}(E)
/
X(E)
\over\displaystyle
\left[
\phi_{\nu_{e}}(E)
/
X(E)
\right]_{\mathrm{max}}
}
\;.
\label{E819}
\end{equation}
Furthermore,
from the NC event rate
we obtain the following upper bound:
\begin{equation}
\left\langle
\mathrm{P}_{\nu_{e}\to\nu_{e}}
\right\rangle_{\mathrm{NC}}
\le
{\displaystyle
N^{\mathrm{NC}}
\over\displaystyle
X_{{\nu}d}^{\mathrm{NC}}
\,
\Phi_{\mathrm{B}}^{\circ}
}
\;.
\label{E831}
\end{equation}
Additional informations
about the probability
$ \mathrm{P}_{\nu_{e}\to\nu_{e}} $
can be obtained from the measurement of the ES event rate
and ES recoil spectrum.
Let us determine the average value:
\begin{equation}
\left\langle
\mathrm{P}_{\nu_{e}\to\nu_{e}}
\right\rangle_{\nu_{e}e}
\equiv
{\displaystyle
1
\over\displaystyle
X_{\nu_{e}e}
}
\int_{E_{\mathrm{th}}^{\mathrm{ES}}}
\sigma_{\nu_{e}e}(E)
\,
X(E)
\,
\mathrm{P}_{\nu_{e}\to\nu_{e}}(E)
\,
\mathrm{d} E
=
{\displaystyle
N^{\mathrm{ES}}
\over\displaystyle
X_{\nu_{e}e}
\,
\Phi_{\mathrm{B}}
}
\;,
\label{E815}
\end{equation}
We obtain
\begin{equation}
\left\langle
\mathrm{P}_{\nu_{e}\to\nu_{e}}
\right\rangle_{\nu_{e}e}
\le
{\displaystyle
N^{\mathrm{ES}}
\over\displaystyle
X_{\nu_{e}e}
\,
\Phi_{\mathrm{B}}^{\circ}
}
\;,
\label{E813}
\end{equation}
Finally, we have
\begin{equation}
\left\langle
\mathrm{P}_{\nu_{e}\to\nu_{e}}
\right\rangle_{\nu_{e}e;\mathrm{T}}
\equiv
{\displaystyle
1
\over\displaystyle
X_{\nu_{e}e}(\mathrm{T})
}
\int_{E_{\mathrm{m}}(\mathrm{T})}
{\displaystyle
\mathrm{d} \sigma_{\nu_{e}e}
\over\displaystyle
\mathrm{d} \mathrm{T}
}
(E,\mathrm{T})
\,
X(E)
\,
\mathrm{P}_{\nu_{e}\to\nu_{e}}(E)
\,
\mathrm{d} E
=
{\displaystyle
n^{\mathrm{ES}}(\mathrm{T})
\over\displaystyle
X_{\nu_{e}e}(\mathrm{T})
\,
\Phi_{\mathrm{B}}
}
\;.
\label{E832}
\end{equation}
{}From Eq.(\ref{E832}) it follows that
\begin{equation}
\left\langle
\mathrm{P}_{\nu_{e}\to\nu_{e}}
\right\rangle_{\nu_{e}e;\mathrm{T}}
\le
{\displaystyle
n^{\mathrm{ES}}(\mathrm{T})
\over\displaystyle
X_{\nu_{e}e}(\mathrm{T})
\,
\Phi_{\mathrm{B}}^{\circ}
}
\;.
\label{E814}
\end{equation}

Thus,
in the case under consideration
we can obtain informations
about the probability of $\nu_e$'s to survive
not only from the CC data, but also from the ES and NC data.

Finally,
we will present the results
of a calculation of some of the lower bounds
for the
$ \nu_e \to \nu_s $
transition probability
in a simple model with
$ \nu_e $--$ \nu_{s} $
mixing which,
as was mentioned in the introduction,
can explain the existing solar neutrino data.
The values of the parameters of the model are
given in Eq.(\ref{E:SMASTE}).
In our calculations
we assumed the SSM $^8\mathrm{B}$ neutrino flux
in order to estimate
the number of CC and NC
events in the SNO experiment
and the number of ES events in the S-K experiment
after one year of data taking.
{}From these numbers
we estimated the statistical accuracy
with which the calculated quantities
will be measured.
We obtained
$ R^{\mathrm{ES}} = 0.00 \pm 0.02 $
and
$ R^{\mathrm{NC}} \le 0.03 \pm 0.03 $.
Therefore,
it will be possible to establish
the absence of
$ \nu_{e} \to \nu_{\mu(\tau)} $
transitions with a good accuracy.
For the lower bound of
the total flux of $^8\mathrm{B}$ neutrinos,
from Eq.(\ref{E560})
we found
$
\Phi_{\mathrm{B}}
\ge
( 3.0 \pm 0.2 ) \times 10^{6} \, \mathrm{cm}^{-2} \, \mathrm{sec}^{-1}
$.
Using this result,
we obtained
$ \displaystyle
\left\langle
\mathrm{P}_{\nu_{e}\to\nu_{s}}
\right\rangle_{\mathrm{NC}}
\ge
0.24 \pm 0.04
$
and
$ \displaystyle
\left\langle
\mathrm{P}_{\nu_{e}\to\nu_{s}}
\right\rangle_{\mathrm{CC}}
\ge
0.22 \pm 0.04
$.

\begin{figure}[p]
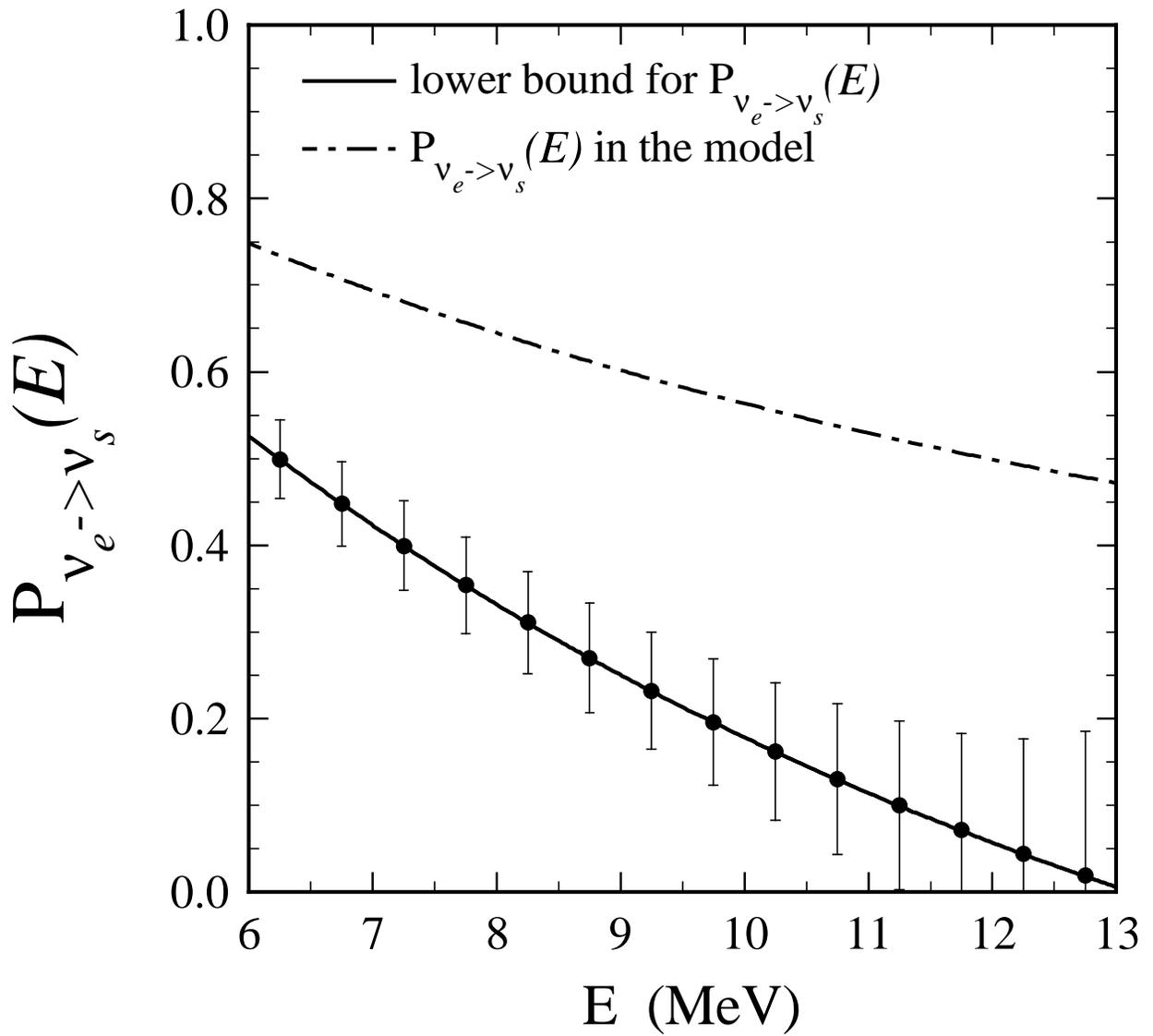

\protect\caption{
Plot of
the lower bound for the probability
$ \displaystyle
\mathrm{P}_{\nu_{e}\to\nu_{s}}(E)
$
as a function of the neutrino energy $E$
calculated in a model with
$ \nu_e $--$ \nu_{s} $
mixing (solid line).
The 1$\sigma$ error-bars are our estimation of
the statistical accuracy with which
this lower bound will be determined
after one year of data taking.
The dash-dotted line represent the value of
$ \displaystyle
\mathrm{P}_{\nu_{e}\to\nu_{s}}(E)
$
calculated in the model.
}
\label{F:PSTE}
\end{figure}

The results of our calculations
for the lower bounds
(\ref{E813}), (\ref{E814}) and (\ref{E819})
are presented in
Fig.\ref{F:PES}, Fig.\ref{F:PEST} and Fig.\ref{F:PSTE}.
The solid line in Fig.\ref{F:PES}
represent the lower bound for
$
\left\langle
\mathrm{P}_{\nu_{e}\to\nu_{s}}
\right\rangle_{\nu_{e}e}
$
as a function of the kinetic energy threshold
$ \mathrm{T}_{\mathrm{th}} $
of the recoil electrons in the ES process.
The dotted lines represent
the 1$\sigma$ statistical
errors after one year of data taking.
The dash-dotted line represent
the value of
$
\left\langle
\mathrm{P}_{\nu_{e}\to\nu_{s}}
\right\rangle_{\nu_{e}e}
$
in the model.
It can be seen from Fig.\ref{F:PES}
that
the lower bound for
$
\left\langle
\mathrm{P}_{\nu_{e}\to\nu_{s}}
\right\rangle_{\nu_{e}e}
$
is bigger for a lower energy threshold.
The solid lines in
Fig.\ref{F:PEST} and Fig.\ref{F:PSTE}
represent the results of the calculations of
the lower bounds for
$
\left\langle
\mathrm{P}_{\nu_{e}\to\nu_{s}}
\right\rangle_{\nu_{e}e;\mathrm{T}}
$
and
$
\mathrm{P}_{\nu_{e}\to\nu_{s}}(E)
$,
respectively.
The error-bars
represent our estimation of
the 1$\sigma$ statistical accuracy
with which these lower bounds could be determined
after one year of data taking.
The dash-dotted lines represent
the behaviour of the probabilities
$
\left\langle
\mathrm{P}_{\nu_{e}\to\nu_{s}}
\right\rangle_{\nu_{e}e;\mathrm{T}}
$
and
$
\mathrm{P}_{\nu_{e}\to\nu_{s}}(E)
$
in the model.

Our estimations illustrate that
the future SNO and S-K experiments
have reasonable possibilities to reveal
transitions of solar neutrinos into sterile states.

\section{Conclusions}

A discovery of transitions
of solar $\nu_e$'s into sterile states
would be a direct discovery of
new physics beyond the Standard Model.
We showed here that the future
solar neutrino experiments SNO and S-K,
in which neutrinos from $^8\mathrm{B}$ decay
will be detected,
may allow to reveal the presence of
$ \nu_e \to \nu_{s} $ transitions
and to obtain lower bounds
for averaged values of the probability
$ \mathrm{P}_{\nu_{e}\to\nu_{s}} $.
We showed also that
it will be possible to obtain
from the data of
the SNO and S-K experiments
a model independent lower bound
for the total flux of $^8\mathrm{B}$ neutrinos
in the general case of transitions of solar $\nu_e$'s
into active as well as into sterile neutrinos.

In the derivation of the expressions
of the lower bounds for
the total flux of $^8\mathrm{B}$ neutrinos
and for the averages of the probability
$ \mathrm{P}_{\nu_{e}\to\nu_{s}} $
we took into account that
in the SNO experiment
solar neutrinos will be detected through the observation
of the CC and ES processes (\ref{E:CC}) and (\ref{E:ES})
as well as the pure NC process (\ref{E:NC}).
In the ICARUS experiment
\cite{B:ICARUS}
the high energy $^8\mathrm{B}$ neutrinos
will be detected through the observation of the CC process
$ \nu_e + ^{40}\mathrm{Ar} \to e^{-} + ^{40}\mathrm{K}^{*} $
and the ES process (\ref{E:ES}).
A detailed investigation of the CC
electron spectrum will be carried out.
The ICARUS experiment
can also reveal the presence of sterile neutrinos
in the solar neutrino flux on the earth
in a model independent way.
In the case of the ICARUS experiment,
a lower bound for the total flux of $^8\mathrm{B}$ neutrinos
can be obtained
from the relations
(\ref{E560}), (\ref{E809}), (\ref{E802}),
(\ref{E804}), (\ref{E524}) and (\ref{E539})
and
lower bounds for
$ \displaystyle
\left\langle
\mathrm{P}_{\nu_{e}\to\nu_{s}}
\right\rangle_{\mathrm{ES}}
$
and
$ \displaystyle
\left\langle
\mathrm{P}_{\nu_{e}\to\nu_{s}}
\right\rangle_{\mathrm{ES};\mathrm{T}}
$
can be obtained from
the relations (\ref{E807}) and (\ref{E808}).
It is necessary, however,
to stress that the contribution
of the total probability
$ \displaystyle
\sum_{\ell=e,\mu,\tau} \mathrm{P}_{\nu_{e}\to\nu_{\ell}} $
to the ES event rate is suppressed by the
smallness of the ratio
$ \sigma_{\nu_{\mu}e} / \sigma_{\nu_{e}e} $.
In order to obtain from the ES and CC
processes informations about the
$ \nu_e \to \nu_{s} $ transitions
it is necessary to have high statistics
CC and ES data.
The ICARUS experiment is assumed to
have this characteristic.


\begin{figure}[p]
\protect\caption{
Plot of the lower bound for
$ \displaystyle
\left\langle
\mathrm{P}_{\nu_{e}\to\nu_{s}}
\right\rangle_{\nu_{e}e}
$
as a function of the kinetic energy threshold
$ \mathrm{T}_{\mathrm{th}} $
of the recoil electrons in the ES process
calculated in a model with
$ \nu_e $--$ \nu_{s} $
mixing (solid line).
The dotted lines are
our estimation of the 1$\sigma$ statistical
accuracy with which
this lower bound
will be determined
after one year of data taking.
The dash-dotted line represent
the value of
$
\left\langle
\mathrm{P}_{\nu_{e}\to\nu_{s}}
\right\rangle_{\nu_{e}e}
$
calculated in the model.
}
\label{F:PES}
\end{figure}

\begin{figure}[p]
\protect\caption{
Plot of the lower bound for
$ \displaystyle
\left\langle
\mathrm{P}_{\nu_{e}\to\nu_{s}}
\right\rangle_{\nu_{e}e;\mathrm{T}}
$
as a function of the kinetic energy
$ \mathrm{T} $
of the recoil electrons in the ES process
calculated in a model with
$ \nu_e $--$ \nu_{s} $
mixing (solid line).
The 1$\sigma$ error-bars
are our estimation of the statistical
accuracy with which this lower bound
will be determined
after one year of data taking.
The dash-dotted line represent
the value of
$ \displaystyle
\left\langle
\mathrm{P}_{\nu_{e}\to\nu_{s}}
\right\rangle_{\nu_{e}e;\mathrm{T}}
$
calculated in the model.
}
\label{F:PEST}
\end{figure}

\end{document}